\documentclass[conference]{IEEEtran}
\IEEEoverridecommandlockouts
\def\ninept{\def\baselinestretch{.95}\let\normalsize\small\normalsize}

\usepackage{amsmath,amsfonts}
\usepackage{algorithmic}
\usepackage{algorithm}
\usepackage{array}
\usepackage[caption=false,font=normalsize,labelfont=sf,textfont=sf]{subfig}
\usepackage{textcomp}
\usepackage{stfloats}
\usepackage{url}
\usepackage{hyperref}
\usepackage{verbatim}
\usepackage{graphicx}
\usepackage{cite}
\usepackage{multirow,enumitem}
\usepackage[mathscr]{eucal}
\usepackage{booktabs}
\usepackage{footmisc}
\usepackage{xcolor}
\usepackage{bbding}
\usepackage[skins]{tcolorbox}
\def\BibTeX{{\rm B\kern-.05em{\sc i\kern-.025em b}\kern-.08em
    T\kern-.1667em\lower.7ex\hbox{E}\kern-.125emX}}
\ninept
\newcommand{\system}{MERITS-L}

\begin{document}

\title{LLM supervised Pre-training for Multimodal Emotion Recognition in Conversations
\thanks{This work was carried out with research grants from British Telecom and the Prime Minister's Research Fellowship.}
}

\author{\IEEEauthorblockN{Soumya Dutta}
\IEEEauthorblockA{\textit{LEAP Lab, Electrical Engineering} \\
\textit{Indian Institute of Science Bangalore}\\
Bangalore, India \\
soumyadutta@iisc.ac.in}
\and
\IEEEauthorblockN{Sriram Ganapathy}
\IEEEauthorblockA{\textit{LEAP Lab, Electrical Engineering} \\
\textit{Indian Institute of Science Bangalore}\\
Bangalore, India \\
sriramg@iisc.ac.in}
}

\maketitle

\begin{abstract}
Emotion recognition in conversations (ERC) is challenging due to the multimodal nature of the emotion expression. In this paper, we propose to pretrain a text-based recognition model from unsupervised speech transcripts with LLM guidance. These transcriptions are obtained from a raw speech dataset with a pre-trained ASR system. A text LLM model is queried to provide pseudo-labels for these transcripts, and these pseudo-labeled transcripts are subsequently used for learning an utterance level text-based emotion recognition model. 
We use the utterance level text embeddings for emotion recognition in conversations along with speech embeddings obtained from a recently proposed pre-trained model. A hierarchical way of training the speech-text model is proposed, keeping in mind the conversational nature of the dataset. We perform experiments on three established datasets, namely, IEMOCAP, MELD, and CMU- MOSI, where we illustrate that the proposed model improves over other benchmarks and achieves state-of-the-art results on two out of these three datasets.
\end{abstract}


\section{Introduction}
Emotion Recognition in Conversation (ERC) focuses on detecting emotions conveyed through multiple modalities during social conversational interactions, which is essential for natural human communication.  Developing artificial systems that have improved emotional understanding and intelligence is a vital design step in conversational agents~\cite{pantic2005affective}, social media analytics tools~\cite{gaind2019emotion}, customer service centers~\cite{li2019acoustic},  mental health monitoring platforms~\cite{ghosh2019emokey}, and wearable systems~\cite{park2020k}. ERC enables these technologies to better adapt to human emotions, enhancing user experiences.

Emotion recognition in conversational data is challenging due to overlapping speakers, short- and long-term dependencies~\cite{poria2019emotion}, short-speaker turns, reverberation and background noise. Emotions are often multimodal, conveyed through various modes such as facial expressions~\cite{tarnowski2017emotion}, vocal cues~\cite{scherer2003vocal}, gestures~\cite{navarretta2012individuality}, and physiological signals~\cite{knapp2011physiological}. To address these complexities, multimodal approaches are often preferred~\cite{poria2017review}. This paper focuses on joint emotion recognition from audio and text, using the strengths of both modalities to enhance accuracy in detecting emotions in conversations.


The initial methods for Speech Emotion Recognition (SER) relied on handcrafted acoustic features like pitch~\cite{lieberman1962some}, energy, and speaking rate~\cite{petrushin1999emotion}. The introduction of deep learning techniques, including CNNs~\cite{yenigalla2018speech}, LSTMs~\cite{hsiao2018effective}, and transformers~\cite{dutta2022multimodal}, significantly improved the SER performance. Recently, self-supervised learning models like wav2vec2.0~\cite{baevski2020wav2vec}, HuBERT~\cite{hsu2021hubert}, and WavLM~\cite{chen2022wavlm} have shown promise in recognition of emotions across multiple datasets and tasks. Large language models (LLMs) have also been explored for SER~\cite{tang2023salmonn,hu2024wavllm}, though they demand significant computational resources.

In parallel, text-based emotion recognition (commonly known as sentiment analysis) initially relied on rule-based methods that linked specific words to emotions~\cite{sebastiani2006sentiwordnet,taboada2011lexicon}. With the rise of deep learning, sentiment analysis progressed to CNNs~\cite{kim-2014-convolutional}, RNNs~\cite{irsoy2014opinion}, and transformer architectures such as BERT~\cite{devlin2019bert, hoang2019aspect} and RoBERTa~\cite{liu2019roberta,raffel2020exploring}. More recently, large language models (LLM) are seen as excellent tools for sentiment analysis~\cite{zhang2023sentiment}. 

In order to enhance emotion recognition in conversations, several works have also designed  multi-modal fusion techniques combining audio and text data, using models like transformers~\cite{dutta2023hcam}, graph neural networks~\cite{ghosal2019dialoguegcn}, and capsule networks~\cite{li-etal-2022-emocaps}.

 

In this paper, we propose a pre-training methodology   through a multi-modal approach that leverages both text and speech representations. Specifically, we introduce a strategy to improve emotion classification from text by leveraging unsupervised speech data with large-scale language models (LLMs). To this end, we first utilize a pre-trained ASR based on Whisper-large model \cite{radford2023robust} to  transcribe speech. The ``noisy'' speech transcripts are labeled with an LLM to automatically generate pseudo-labels of speech sentiments. These labelled text-transcripts are then used to fine-tune a RoBERTa text encoder model \cite{liu2019roberta} for sentiment classification, allowing it to capture nuanced emotional patterns in textual data. 
We show substantial benefits from this unsupervised pre-training of the text-based model. 

On the speech side, we extract features using the recently proposed CARE model~\cite{dutta2024leveraging}. The CARE model is designed to generate high-quality embeddings that encapsulate both content and acoustic information from speech utterances. Using the speech and text embeddings derived at utterance level, we train a bi-directional gated recurrent unit (GRU) based model to assimilate the information across the entire conversation.  The conversation-level embeddings from uni-modal speech-text models  are integrated into our proposed multi-modal architecture for conversational emotion recognition. We propose a hierarchical fusion mechanism with a cross-attention-based network~\cite{lu2019vilbert} to  enable the interaction between the modalities, ensuring an fusion of the emotional information present in both speech and text. We call our method \system{}  (\textbf{M}ultimodal  \textbf{E}motion \textbf{R}ecognition \textbf{I}n \textbf{S}peech and \textbf{T}ext with \textbf{L}LM guidance). 

The experiments are performed on three established datasets, namely IEMOCAP~\cite{busso2008iemocap}, MELD~\cite{poria2019meld} and CMU-MOSI~\cite{zadeh2016mosi}. 
The key contributions from this work are:
\begin{itemize}[leftmargin=*]
    \item We propose a pre-training methodology for improving text emotion recognition. We make use of an unsupervised speech   corpus and a large language model (LLM) for this purpose. 
    \item We propose a hierarchical approach for the multi-modal emotion recognition, where the information is first processed at the utterance level in each of the modalities, followed by inter-utterance conversation modeling. Subsequently, multi-modal processing with co-attention is designed. 
    \item We evaluate the proposed \system{} model on three benchmark datasets and achieve state-of-the-art results for two out of these  three datasets. 
    %
\end{itemize}\par
\section{Related Work}
\noindent \textbf{LLMs for Text Sentiment Analysis}: The capabilities of large language models (LLMs) have been the focus of recent research efforts~\cite{wang2023chatgpt,zhong2023can,zhang2023sentiment}. Zhong et al.\cite{zhong2023can} demonstrated that ChatGPT\footnote{\url{https://chatgpt.com}} achieves performance comparable to fine-tuned BERT models. Zhang et al.\cite{zhang2023sentiment}, however, provided a more comprehensive evaluation across various sentiment analysis tasks. Their findings reveal that while LLMs under-perform in fine-grained sentiment analysis, they exhibit promising zero-shot capabilities in simpler tasks, such as binary sentiment classification. In this work, we leverage the ability of LLMs to coarsely annotate large corpora of emotional speech  transcripts. \\
\par
\noindent \textbf{Emotion Recognition in Conversations}:  Recent methods that achieve strong performance on benchmark datasets often incorporate speaker identity~\cite{song2022emotionflow,hu2023supervised,liu2023hierarchical, yu2024emotion}. For instance, Hu et al.\cite{hu2023supervised} introduced  a supervised contrastive loss, where utterances with the same emotion and speaker are treated as positive samples in a contrastive learning framework. Yu et al.\cite{yu2024emotion} appended speaker embeddings to the utterance representations and utilize a language model, such as RoBERTa~\cite{liu2019roberta}, to predict emotions via masked language modeling. In contrast, our work does not access speaker labels for any utterance in the conversation.\par

\section{Method}
\begin{figure*}
    \centering
    \includegraphics[width=0.9\textwidth,trim={0cm 2.5cm 3.5cm 2.1cm},clip]{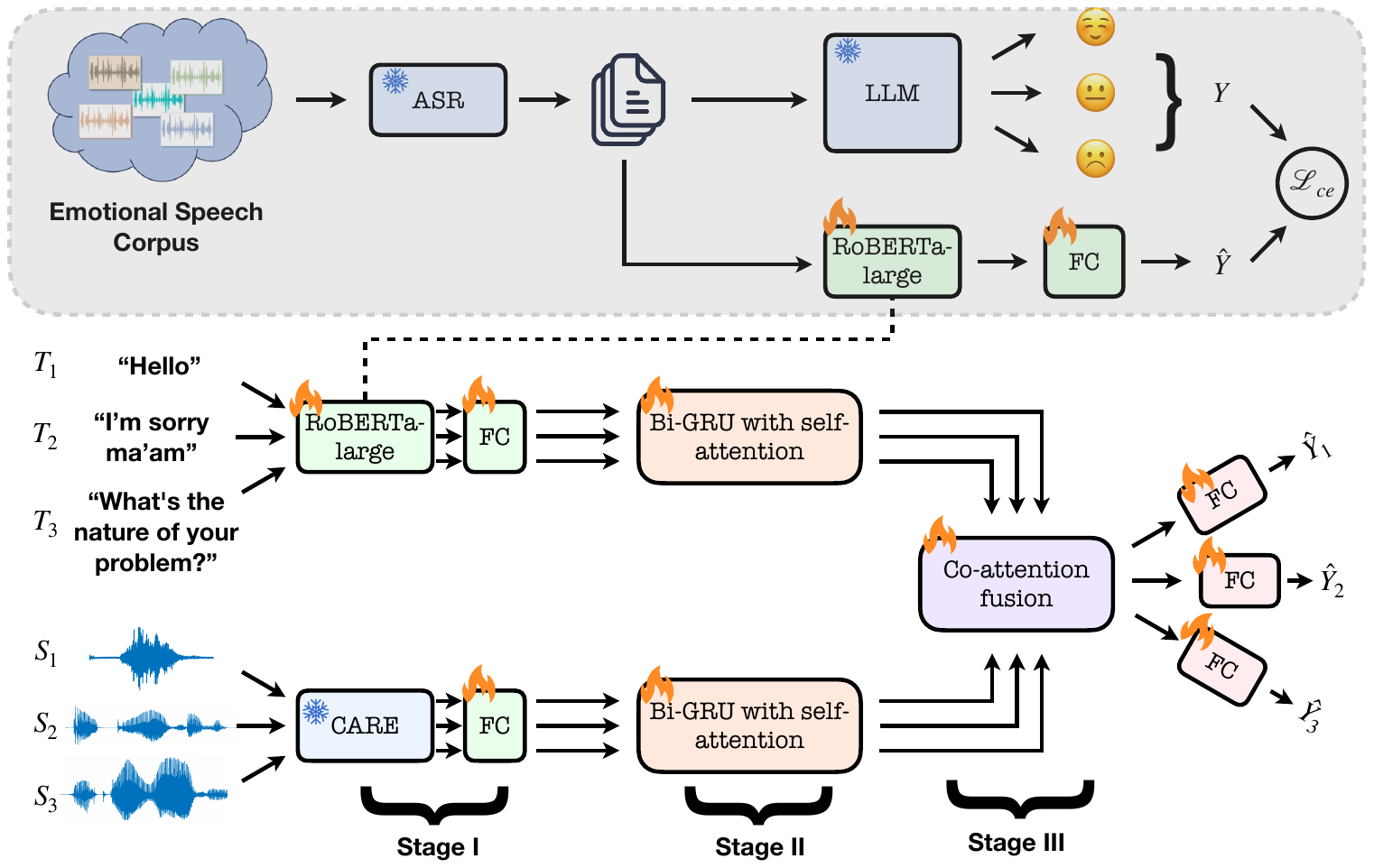}
    \caption{Block diagram of the proposed model. The pre-training stage is shown in the grey box at the top. An ASR system is used to generate the transcripts for the pre-training data which are annotated by a large language model (LLM) as positive, negative or neutral sentiment. These ``silver'' labels with the text transcripts form the supervised training  dataset  for RoBERTa-large model.   A frozen CARE model~\cite{dutta2024leveraging} is used for extracting  audio embeddings. Both the text and speech embeddings thus use only unsupervised data.  The \system{} model is trained in three stages (denoted as Stage I, II and III in the diagram), wherein the models trained in a particular stage are kept frozen for subsequent stages.  }
    \label{fig:entire model}
\end{figure*}
\subsection{Background}
\subsubsection{CARE} In our recently proposed CARE model~\cite{dutta2024leveraging}, speech is processed by two encoders - one focusing on the semantic aspect of speech by aligning with the mean-pooled RoBERTa representation of corresponding ASR transcripts, while the other is trained to predict low-level descriptors of speech provided by the PASE+ model~\cite{ravanelli2020multi}. The CARE embeddings are seen to perform better than most other base-sized models in the SUPERB~\cite{yang2021superb} style evaluation. In this paper, we utilize the CARE embeddings from speech utterances and train conversational models along with speech-text fusion.
\subsubsection{RoBERTa} One of the significant contributions in creating a large scale language model was proposed by Devlin et al.~\cite{devlin2019bert}. This architecture, called bidirectional encoder representations from transformer (BERT), was trained with two objectives on a corpus of textual data, namely, predicting words masked out in a sentence and to predict whether two sentences semantically follow each other (referred to as next sentence prediction). Liu et. al \cite{liu2019roberta} trained this architecture on a larger corpus of textual data without the next sentence prediction task. This pre-trained model is known as robust optimized BERT approach (RoBERTa). We use the pre-trained large version of this model with $24$ transformer layers as the text encoder.

\subsection{Proposed \system{} model} The block diagram of the proposed model is shown in Fig.~\ref{fig:entire model}.

\subsubsection{Problem Description} Given a set of utterances $U$ and set of emotion labels $Y$, a conversation consisting of $K$ utterances is denoted by $[(u_1,y_1),(u_2,y_2),\dots,(u_K,y_K)]$, where $y_j\in Y$ is the emotion of utterance $u_j$ in the conversation. Specifically, the task of ERC is achieved by using the speech and text modalities in our case, which means $u_k=\{S_k, T_k\}$, where $S_k$ and $T_k$ refer to the speech and the text transcript associated with the utterance $u_k$. The objective of ERC is to predict the emotion label $y_k$ of each utterance $u_k$.

\subsubsection{LLM guided text pre-training} The text transcripts from the emotional speech corpus are first extracted from an ASR system (Whisper-large-v3 \cite{radford2023robust}). Generally, the word error rates on emotional speech are higher than neutral speech~\cite{li2023asr}.  With the ASR generated transcripts of the speech corpus, a large language model (LLM) is prompted to annotate the transcript as three classes, ``positive'', ``negative'' or ``neutral''.  The pre-trained RoBERTa-large model is fine-tuned to predict the pseudo-classes (from the LLM predictions) and the resultant model is used subsequently as a text feature extractor at utterance-level. This model is referred to as RoBERTa-FT in the subsequent sections of the paper.

\subsection{Training} The training methodology for \system{} is   performed in stages as mentioned below:\par
\noindent \textbf{Stage I}: All utterances, $U$, are collated and used to train text sentiment analysis and speech emotion recognition models for each dataset. While the RoBERTa-FT model is fine-tuned for each downstream dataset, small light weight networks with frozen CARE embeddings as input are trained for the speech modality for every dataset. This training stage aims to classify the text transcript ($T_k$) and the speech signal ($S_k$) into the correct emotion category ($y_k$).
The final layer embeddings for each utterance ($T_k^{1}$ and $S_k^{1}$ for the text transcripts and speech signal   respectively) are used for the next stage of training.\par
\noindent \textbf{Stage II}: This stage introduces the conversational nature of the data in the modeling framework. The text features for the utterances in a conversation from the previous stage, denoted by $(T_1^1, T_2^1, \dots, T_K^1)$ for a conversation having $K$ utterances are   processed by a bidirectional gated recurrent network (Bi-GRU) with self-attention over the conversational context. This stage encourages the model to predict the emotion class of the utterance $u_k$, keeping the entire conversation  as a part of the  context. A similar modeling exercise is done for the speech modality as well. Similar to the previous stage, the features from the Bi-GRU with self-attention blocks are used for the final stage of training. These features are denoted by $T_k^{2}$ and $S_k^{2}$ for the text and speech modality, of utterance $u_k$,  respectively. \par
\begin{figure}
    \centering
    \includegraphics[width=\columnwidth,trim={2cm 4.5cm 0cm 2cm},clip]{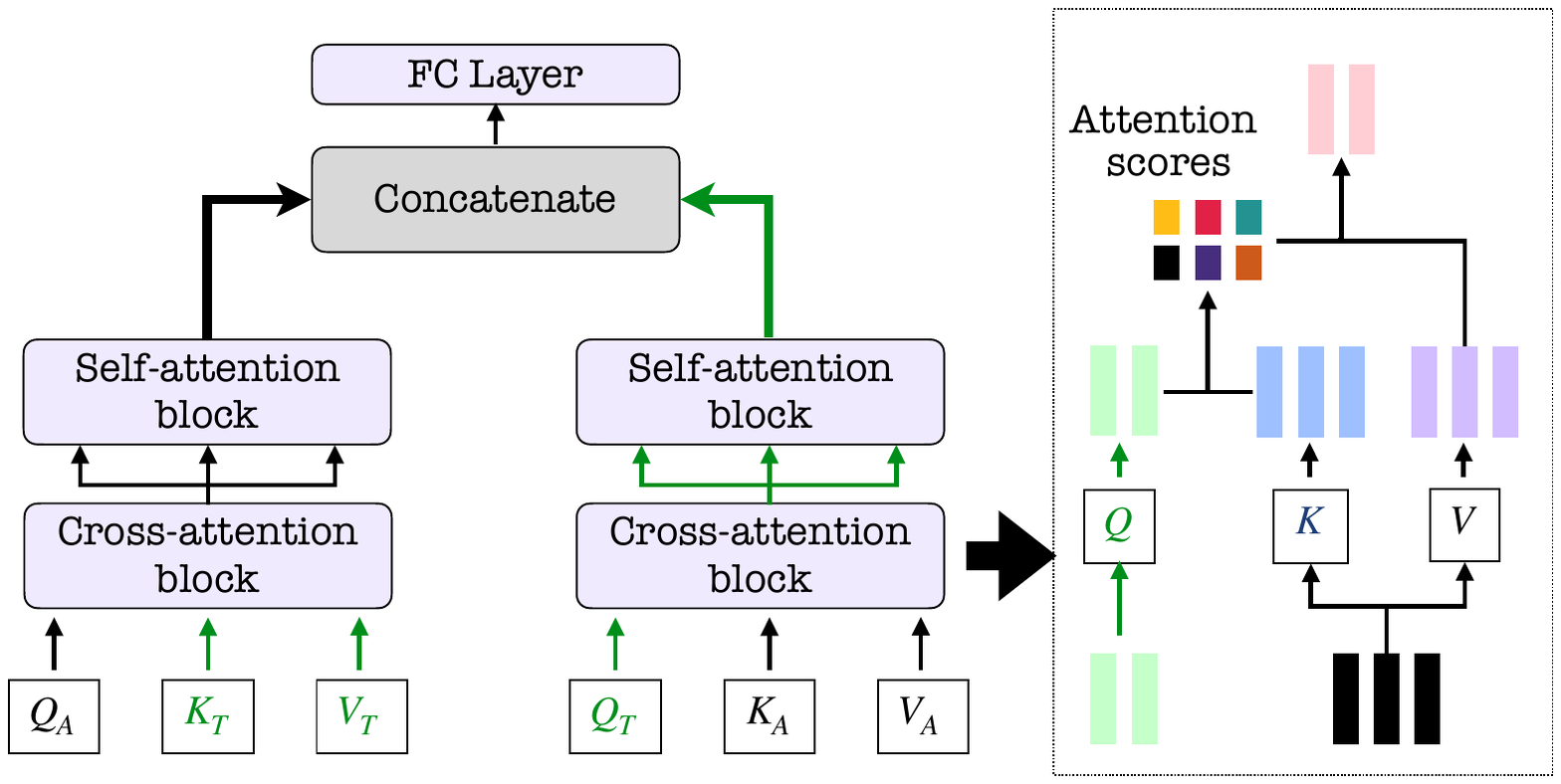}
    \caption{The co-attention network used in the proposed model. It consists of two sub-blocks - the cross-attention and the self-attention blocks.}
    \label{fig:attention}
\end{figure}
\noindent \textbf{Stage III}: Notably, the previous stages included training the two modalities separately. In order to align the two modalities in a more effective way for emotion recognition, they are combined in this stage. We implement a co-attention fusion strategy for the two modalities as outlined in Fig.~\ref{fig:attention}. The query-key-value sequences for the two modalities are the features obtained after Stage II of training. E.g. in Fig.~\ref{fig:attention}, $Q_A$ refers to the query sequence of the speech modality and is denoted as $(S_1^2, S_2^2, \dots, S_K^2)$ for the conversation $C$.

\section{Experiments and Results}
\subsection{Datasets}
\noindent \textbf{Pre-training}:The MSP-PODCAST corpus~\cite{lotfian2017building} is used for the task of pre-training. A total of $149,307$ speech-turns amounting to   $230$ hours of emotional speech data is used. Out of the total number of samples, $80\%$ of the data is randomly chosen as the training set while the remaining $20\%$ serves as the validation set. The Whisper-large-v3 model\footnote{\url{https://huggingface.co/openai/whisper-large-v3}} is used for generating the transcripts. These transcripts are annotated using the \texttt{GPT-3.5 Turbo}\footnote{\url{https://platform.openai.com/docs/models/gpt-3-5-turbo}} model.

\noindent \textbf{ERC datasets}: Three datasets are used for evaluating \system{} on the ERC task - IEMOCAP~\cite{busso2008iemocap}, MELD~\cite{poria2019meld} and CMU-MOSI~\cite{zadeh2016mosi}.

\noindent \textbf{IEMOCAP dataset:} The IEMOCAP dataset consists of $151$ video recordings split into 5 sessions. Keeping in line with previous works, we do a four-way classification task where we consider ``angry'', ``happy'', ``sad'', ``neutral'' and ``excited'' categories (with excited and happy categories merged). We have a total of $5531$ utterances from the four emotion labels. We consider session $5$ for testing purposes. We choose session $1$ for validating our models and sessions $2-4$ for training. 

\noindent \textbf{MELD dataset:} The MELD dataset is a multi-party dataset created from video clippings of the popular TV show, ``Friends''. The training data consists of $9988$ utterances, validation data consists of $1108$ utterances and  test data consists of $2610$ utterances.
A seven way classification task is performed on this dataset, with each utterance being labeled as one of the $7$ emotions - ``angry'', ``sad'', ``joy'', ``neutral'', ``fear'', ``surprise'' or ``disgust''. 

\noindent \textbf{CMU-MOSI dataset:} The CMU-MOSI dataset has a total of $93$ monologues divided into $2199$ utterances. Each utterance is labeled in the range of $[-3, 3]$. Following previous works, we treat this as a binary classification problem with utterances having sentiment values in the range $[-3, 0)$ being classified as negative sentiment and those with values in the range $[0, 3]$ as positive sentiment. For dataset partitioning, we follow the prior work by Poria et. al~\cite{poria2017context}, where the first  $62$ monologues are used for training and validation while the last $31$ monologues are used for testing. Of the $62$ monologues, we use $49$ for training our model and the rest $13$ for validation.

\subsection{Implementation details}
\label{sec:impl}
Once the transcripts are generated by the Whipser model, the \texttt{GPT-3.5 Turbo} model is prompted as follows: \par
\begin{tcolorbox}[enhanced,width=\columnwidth,center upper,size=fbox,
    fontupper=\small\bfseries,
    colframe=red!50!black,colback=yellow!10]
\texttt{You are a sentiment classification bot. Given the [sentence], classify as positive, negative or neutral sentiment. Please give the sentiment and no extra text as output.}
\end{tcolorbox}
\noindent
The RoBERTa-large model is pre-trained with the LLM generated labels ($3$ classes) for a total of $10$ epochs with a learning rate of $1e$-$4$ and a batch size of $32$. The different stages of \system{} are trained for a total of $50$ epochs with a learning rate of $1e$-$4$ and a batch size of $32$. For all training purposes, the cross-entropy loss with AdamW~\cite{loshchilov2017decoupled} optimizer is used. The weighted F1-score is used as the metric for performance evaluation on the ERC datasets. Note that, we have not used any additional labeled datasets in pre-training as the pre-training framework for speech and text are purely based on self-supervised learning principles from raw data. Further, the downstream datasets are used without any knowledge of speaker meta-data. 
\begin{table}[t!]
\centering
\caption{Results on the datasets in terms of weighted F1-score.}\label{tab:results}

\resizebox{0.9\columnwidth}{!}{%
\begin{tabular}{@{}l|c|c|c@{}}
\toprule
Modality (Training Stage) & IEMOCAP & MELD & CMU-MOSI \\ \midrule
Audio (Stage I) & $66.93$ & $47.61$ & $68.77$ \\
Audio (Stage II) & $77.95$ & $49.32$ & $69.43$ \\ \midrule
Text (Stage I) & $69.84$ & $63.81$ & $85.41$ \\
Text (Stage II) & $83.85$ & $65.24$ & $86.02$ \\ \midrule
Audio+Text (Stage III) & $86.48$ & $66.02$ & $86.81$ \\ \bottomrule
\end{tabular}}
\end{table}

\subsection{Results}
The results for the three stages are shown in Table~\ref{tab:results}. We note that the performance of both the modalities improve with every modeling stage. The introduction of the contextual information is seen to significantly improve the performance of IEMOCAP dataset  (relative improvements of $16.46\%$ and $20.06\%$ for audio and text modality respectively). The performance of the two modalities are seen to be comparable for the IEMOCAP dataset, unlike the other two where text emotion recognition performance is considerably higher than the performance with the audio modality. Finally, the multi-modal fusion is seen to aid all the datasets, achieving relative improvements of $16.28\%$, $2.24\%$ and $5.65\%$ over the best performing modality (after Stage II) for IEMOCAP, MELD and CMU-MOSI respectively. This also shows that the multi-modal fusion strategy is more effective while combining modalities having comparable performance. This is perhaps due to the symmetrical nature of the co-attention based fusion mechanism.
\begin{figure}
    \centering
    \includegraphics[width=0.8\columnwidth,trim={8cm 7.5cm 5cm 5cm},clip]{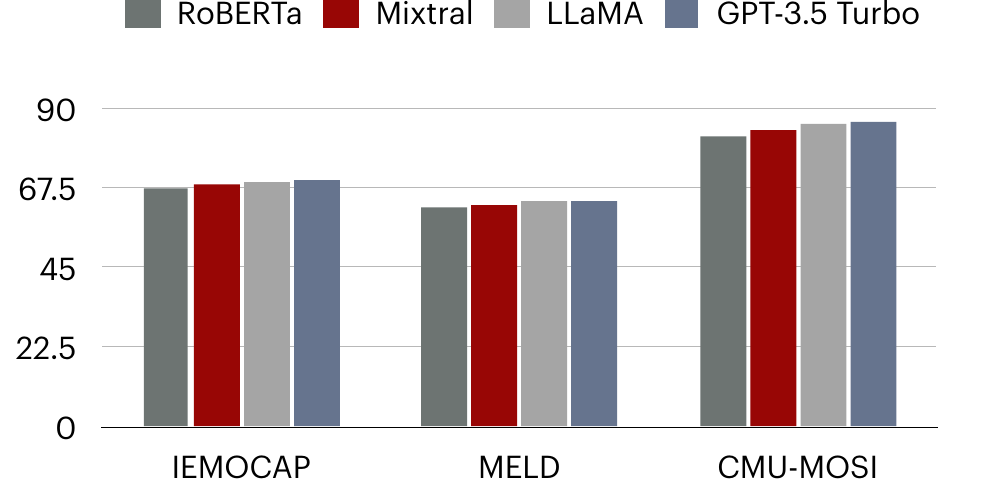}
    \caption{The performance of the   RoBERTa-large models on the different datasets. Different LLMs are used for generating pseudo emotion labels  from speech transcripts. The performance of pre-trained RoBERTa without any supervised fine-tuning is also reported.}
    \label{fig:llm}
\end{figure}
\subsection{Evaluation with different LLMs}
We design an oracle experiment in this regard. Using a similar prompt template,  as described in Sec.~\ref{sec:impl}, we annotate the transcripts into the same three classes using \texttt{Mixtral-8x7B-Instruct-v0.1}\footnote{\url{https://huggingface.co/mistralai/Mixtral-8x7B-Instruct-v0.1}} and \texttt{Llama-3-8b-chat-hf}\footnote{\url{https://huggingface.co/meta-llama/Meta-Llama-3-8B}}. Since the MSP-PODCAST dataset has valence-arousal-dominance values, ranging from $1$ to $7$, we assign a positive label to samples having valence in the range $(5, 7]$, while negative label is assigned to samples having valence value in $[1, 3)$. The rest of the samples are assigned the neutral label. With these labels serving as the ground truth, we notice that \texttt{GPT-3.5 Turbo} achieves a label overlap of $52.98\%$, while \texttt{Llama-3-8b-chat-hf} is comparable with an overlap of $50.91\%$. The performance of \texttt{Mixtral-8x7B-Instruct-v0.1} is the lowest with an overlap of only $44.98\%$. 

While the above benchmarking used oracle emotion labels from the speech dataset, we perform a downstream evaluation using the three choices of LLM. The impact of the different LLM annotation ability is shown in Fig.\ref{fig:llm}, where the performance of the model with the text modality after Stage I training is shown. The performance of the model after Stage I with pre-trained RoBERTa (without any LLM guidance) is also shown for reference. We notice that   the RoBERTa model fine-tuned with labels provided by \texttt{GPT-3.5 Turbo} achieves relative improvements of $8.22\%$, $5.01\%$ and $21.9\%$ for IEMOCAP, MELD and CMU-MOSI respectively over the pre-trained RoBERTa model.   The relative improvements achieved by \texttt{GPT-3.5 Turbo} over \texttt{Mixtral-8x7B-Instruct-v0.1} are $5.4\%$, $3.39\%$ and $12.51\%$ for IEMOCAP, MELD and CMU-MOSI respectively. The highest performance improvement in the CMU-MOSI dataset may be attributed to the  binary classification task in this dataset.
\begin{figure}
    \centering
    \includegraphics[width=0.7\columnwidth,trim={7.5cm 8.5cm 9cm 4cm},clip]{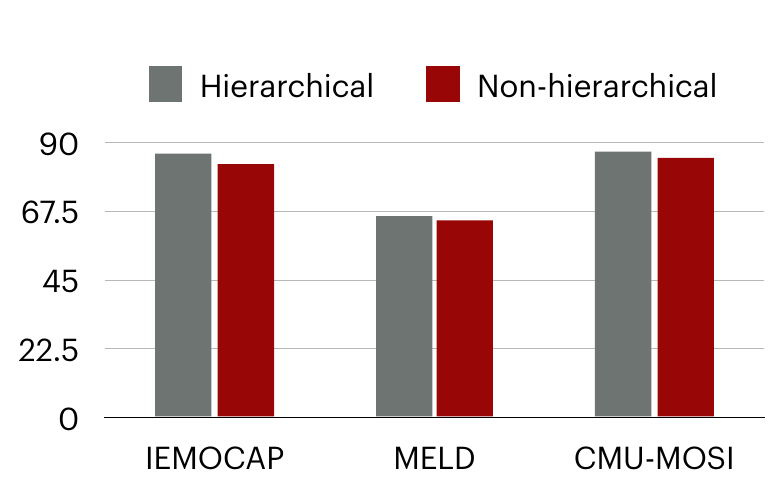}
    \caption{The importance of hierarchical training in \system{}}
    \label{fig:ablation}
\end{figure}
\subsection{Importance of hierarchical training}

In order to understand the impact of hierarchical training, we combine Stages II and III of \system{}. The impact of such a training philosophy is shown in Fig.\ref{fig:ablation}. The impact of this change in the training methodology is found to be the most in the case of IEMOCAP where the performance of \system{} drops from $86.48\%$ to $82.91\%$. A performance drop of around $2\%$ in absolute terms is also noticed for MELD and CMU-MOSI. The benefits from the hierarchical training arises  from the fact that the end-to-end training of the different components of the model often leads to over-fitting as these datasets are relatively small in size. 
\begin{table}[t!]
\centering
\caption{Comparison with different methods. $^*$stands for results with tri-modal system - text, audio and visual modalities. All numbers are weighted F1-scores. The best performing model for each dataset in marked in \textbf{bold}, while \underline{underline} indicates the next best performing model.}\label{tab:compare}

\resizebox{0.9\columnwidth}{!}{%
\begin{tabular}{@{}l|c|c|c@{}}
\toprule
Approaches & \multicolumn{1}{l|}{IEMOCAP} & \multicolumn{1}{l|}{MELD} & \multicolumn{1}{l}{CMU-MOSI} \\ \midrule
LMFN~\cite{mai2019locally} & $82.54$ & - & $80.92$ \\ \midrule
M3ER~\cite{mittal2020m3er} & $82.40$ & - & - \\ \midrule
DialogueTRM~\cite{mao2021dialoguetrm} & - & $63.55$ & - \\ \midrule
SMIN~\cite{lian2022smin} & $\mathbf{87.47}$ & $64.50$ & $81.45$ \\ \midrule
UniMSE~\cite{hu2022unimse} & - & $65.51^{*}$ & $\underline{85.78}$ \\ \midrule
EmoCaps~\cite{li-etal-2022-emocaps} & - & $63.73$ & - \\ \midrule \midrule
\system{} & $\underline{86.48}$ & $\mathbf{66.02}$ & $\mathbf{86.81}$ \\ \bottomrule
\end{tabular}}
\end{table}
\subsection{Comparison with other work}
We compare the performance of \system{} with some recent works in Table~\ref{tab:compare}. The proposed \system{} achieves the best performance when compared with state-of-the-art  models for MELD and CMU-MOSI. For IEMOCAP however, the method by Lian et al.~\cite{lian2022smin} outperforms \system{} by a margin of $1\%$ absolute. Note that, there have been multiple methods like TelME~\cite{yun2024telme} and EACL~\cite{yu2024emotion} which achieve higher performance than \system{} on MELD dataset. However, these methods use information about the speaker identity of each spoken utterance and hence are excluded from the comparison in this work. 
\section{Summary}
In this paper, we first propose a novel way of super-vised pre-training  of text based emotion recognition using LLM guidance. Text from an emotional speech corpus is extracted, following which a text emotion recognition model is trained to classify each transcript using the pseudo-labels. With this text based recognition model as the utterance-level text embedding extractor, we propose \system{}, wherein we develop  the model for emotion recognition in conversations by using the speech and textual modalities. A hierarchical way of training the model is proposed, starting with utterances from a single modality, followed by the contextual modeling at the conversational level and subsequently, the alignment of the two modalities.  Comparison with other state-of-art works indicate the superiority of our method for two out of the three datasets considered in this work.

\bibliographystyle{IEEEtran}
\bibliography{refs}

\end{document}